\documentclass{article}

\usepackage{microtype}
\usepackage{graphicx}
\usepackage{subfigure}
\usepackage{booktabs} 
\usepackage{mdframed}
\usepackage{colortbl}
\usepackage{multirow}
\usepackage{caption}
\usepackage[table, svgnames, dvipsnames]{xcolor}
\usepackage{pifont}
\usepackage{threeparttable}
\usepackage{float}
\usepackage{url}
\usepackage{tabularx} 
\usepackage[dvipsnames]{xcolor}
\usepackage{arydshln} 

\definecolor{lightred}{HTML}{E0115F}     
\definecolor{darkred}{HTML}{8B0000}       

\definecolor{lightblue}{HTML}{BA55D3}     
\definecolor{darkblue}{HTML}{00008B}      

\newcommand{\CHECK}{\textcolor{green}{\ding{51}}} 
\newcommand{\CROSS}{\textcolor{red}{\ding{55}}} 
\newcommand{\dataname}{\textsc{SpatialSoundQA}}
\newcommand{\modelname}{\textsc{Bat}}

\definecolor{mygreen}{HTML}{008000}
\newcommand{\zhisheng}[1]{{\color{orange}{\bf ZZ: }#1}}
\newcommand{\puyuan}[1]{{\color{mygreen}{\bf PP: }#1}}
\newcommand{\eunsol}[1]{{\color{teal}{\bf EC: }#1}}
\newcommand{\david}[1]{{\color{olive}{\bf DH: }#1}}
\newcommand{\xiechen}[1]{{\color{blue}{\bf XC: }#1}}
\newcommand{\ziyang}[1]{{\color{red}{\bf ZY: }#1}}

\renewcommand{\zhisheng}[1]{}
\renewcommand{\puyuan}[1]{}
\renewcommand{\eunsol}[1]{}
\renewcommand{\david}[1]{}
\renewcommand{\xiechen}[1]{}
\renewcommand{\ziyang}[1]{}

\usepackage{hyperref}

\usepackage[accepted]{icml2024}

\usepackage{amsmath}
\usepackage{amssymb}
\usepackage{mathtools}
\usepackage{amsthm}

\usepackage[capitalize,noabbrev]{cleveref}

\theoremstyle{plain}

\theoremstyle{definition}

\theoremstyle{remark}

\usepackage[textsize=tiny]{todonotes}

\mathchardef\mhyphen="2D

\newcommand*{\img}[1]{%
    \raisebox{-.2\baselineskip}{%
        \includegraphics[
        height=1.4\baselineskip,
        width=1.4\baselineskip,
        keepaspectratio,
        ]{#1}%
    }%
}

\icmltitlerunning{\modelname: Learning to Reason about Spatial Sounds with Large Language Models}

\begin{document}

\twocolumn[
\icmltitle{\img{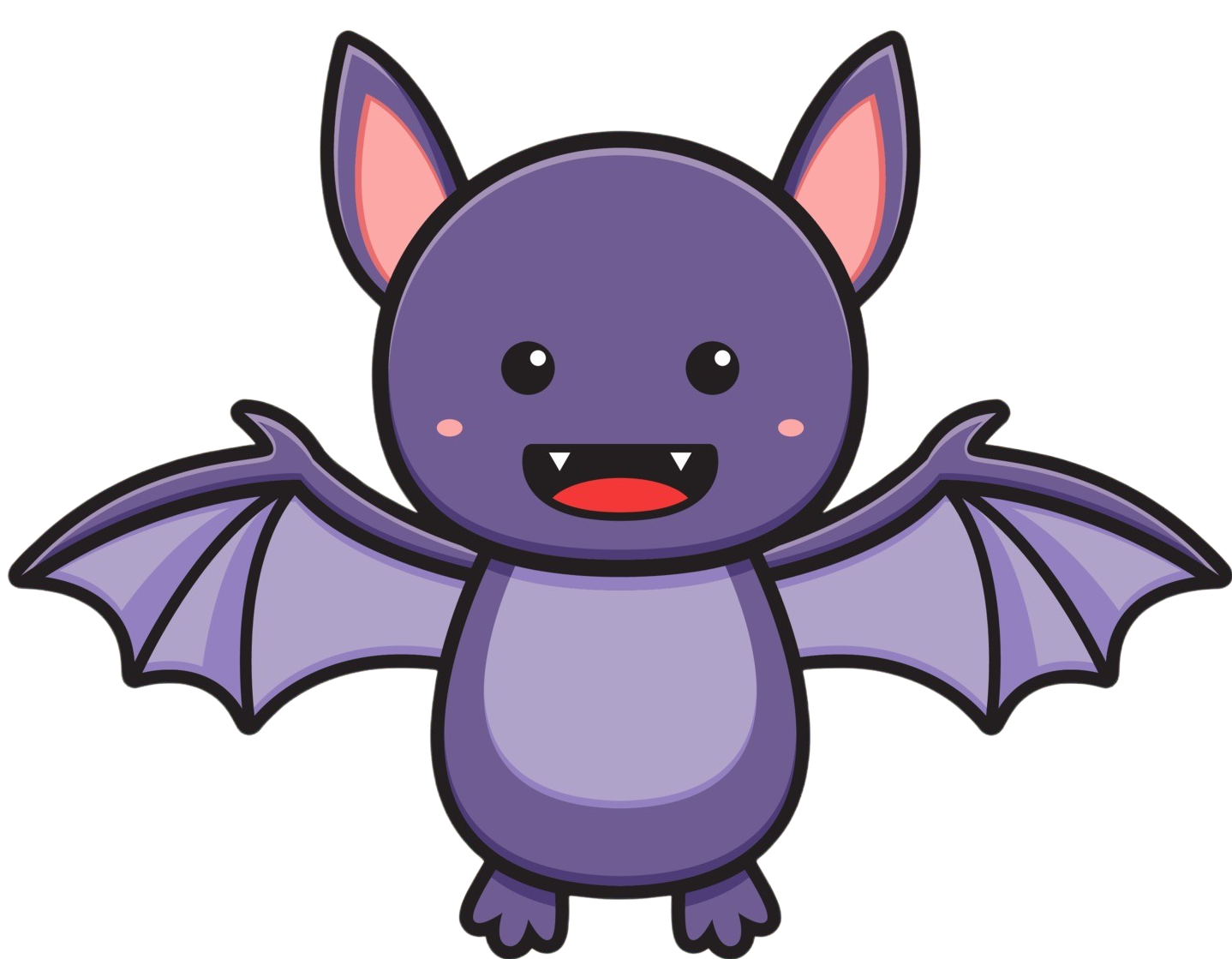}\hspace{-3pt} \modelname: Learning to Reason about Spatial Sounds with Large Language Models}




\icmlsetsymbol{equal}{*}

\begin{icmlauthorlist}
\icmlauthor{Zhisheng Zheng}{utcs,xlance}
\icmlauthor{Puyuan Peng}{utcs}
\icmlauthor{Ziyang Ma}{xlance}
\icmlauthor{Xie Chen}{xlance}
\icmlauthor{Eunsol Choi}{utcs}
\icmlauthor{David Harwath}{utcs}
\end{icmlauthorlist}

\icmlaffiliation{xlance}{Department of Computer Science and Engineering, Shanghai Jiao Tong University, China}
\icmlaffiliation{utcs}{Department of Computer Science, University of Texas at Austin, USA}

\icmlcorrespondingauthor{Zhisheng Zheng}{zszheng@utexas.edu}
\icmlcorrespondingauthor{David Harwath}{harwath@utexas.edu}

\icmlkeywords{Machine Learning, ICML}

\vskip 0.3in
]



\printAffiliationsAndNotice{}  



\begin{abstract}
Spatial sound reasoning is a fundamental human skill, enabling us to navigate and interpret our surroundings based on sound. In this paper we present \modelname, which combines the spatial sound perception ability of a binaural acoustic scene analysis model with the natural language reasoning capabilities of a large language model (LLM) to replicate this innate ability. To address the lack of existing datasets of in-the-wild spatial sounds, we synthesized a binaural audio dataset using AudioSet and SoundSpaces 2.0. Next, we developed \dataname, a spatial sound-based question-answering dataset, offering a range of QA tasks that train \modelname~in various aspects of spatial sound perception and reasoning. The acoustic front end encoder of \modelname~is a novel spatial audio encoder named Spatial Audio Spectrogram Transformer, or \textsc{Spatial-AST}, which by itself achieves strong performance across sound event detection, spatial localization, and distance estimation. By integrating \textsc{Spatial-AST} with LLaMA-2 7B model, \modelname~transcends standard Sound Event Localization and Detection (SELD) tasks, enabling the model to reason about the relationships between the sounds in its environment. Our experiments demonstrate \modelname's superior performance on both spatial sound perception and reasoning, showcasing the immense potential of LLMs in navigating and interpreting complex spatial audio environments. Our demo, dataset, code and model weights are available at: \url{https://zhishengzheng.com/bat}.
\end{abstract}




\section{Introduction}


The evolution of Large Language Models (LLMs) has enabled their application beyond textual data, giving rise to multimodal language models capable of performing image understanding tasks such as visual question answering and image captioning~\cite{li2023blip2, liu2023visual, openai2023gpt4v, peng2023kosmos2}. Other work has integrated sound perception ability into LLMs, enabling them to perform tasks such as audio question answering, speech recognition, speech translation, and speech synthesis~\cite{deshmukh2023pengi, wang2023lauragpt, chu2023qwen, audiolm, gong2023listen}. However, despite these LLM-based models' ability to handle numerous visual and audio tasks with impressive performance, so far none of them can handle in-the-wild spatial audio input. Humans are equipped with binaural hearing ability, allowing us to not only identify the type of sound we are hearing, but also infer how far away the sound is, what direction it is coming from, and whether multiple sound sources at different locations are present at once. Consider scenarios such as locating someone calling from upstairs, hearing your phone ringing from behind a sofa, or sensing footsteps approaching from behind; these situations all demand spatial audio perception and reasoning, a capability not yet present in the aforementioned models.

To bridge this gap, we introduce \modelname, the first spatial audio-based LLM designed to reason about sounds in a 3-D environment. Recognizing the current lack of large-scale datasets for in-the-wild spatial audio and spatial audio-based question answering, we synthesized a large-scale binaural audio dataset using Audioset~\cite{Audioset} clips as sound sources and Soundspaces 2.0~\cite{chen2022soundspaces} for simulating diverse, 3-D, reverberant acoustic environments. In conjunction with this dataset, we developed \dataname, a diverse collection of question answering tasks designed to train and evaluate spatial sound understanding across varying levels of complexity.

The successful training and evaluation of \modelname~hinge on the ability to encode rich spatial audio information accurately. However, existing encoders often specialize in either monaural event detection~\cite{Baade2022MAEASTMA, huang2022masked, chen2022beats, chen2024eat} or are limited to first-order ambisonics~\cite{shimada2021accdoa, Du_NERCSLIP_task3_report, kim2023adyolo}, focusing on a limited range of sound events. Other approaches are solely dedicated to direction-of-arrival (DoA) estimation in specific multi-channel formats~\cite{yang2022enhancing, wang2023fnssl}, lacking the breadth needed for comprehensive spatial audio understanding tasks. 
To overcome these limitations, we developed \textsc{Spatial-AST}, a novel multi-task spatial encoder that is not only capable of sound event detection and spatial localization but also excels in distance perception, thereby providing a comprehensive solution for spatial audio tasks.



In our experiments, we found that for \textsc{Spatial-AST} is both effective and parameter-efficient. Using multi-step training curriculum, the \textsc{Spatial-AST} encoder by itself can achieve strong performance across multiple tasks: a mean Average Precision (mAP) of 50.03\% for audio event classification, a Mean Angular Error (MAE) of 17.94° for direction of arrival estimation, and a Distance Error Rate (DER) within 0.5 meters of the actual location at 32.54\% for distance estimation. By fusing the \textsc{Spatial-AST} encoder to the LLaMA-2~\cite{touvron2023llama2} and adopting an appropriate training curriculum, our \modelname~not only excels in tasks involving perception but also demonstrates spatial reasoning abilities in scenarios with mixed sound sources (e.g. ``Is the stereo system to the left of the barking dog?''), achieving a 76.89\% accuracy in answering spatial sound reasoning questions.

Our key contributions are summarized as follows:
\begin{itemize}
\item We present the first spatial audio-based question answering dataset \dataname, offering a range of 3-D audio understanding tasks from perception to reasoning, which provides a platform for training models to perform spatial sound understanding tasks.
\item We propose \textsc{Spatial-AST}, a binaural spatial audio encoder architecture that jointly performs sound event detection, spatial localization, and distance estimation that achieves strong performance across all three tasks.
\item We introduce \modelname, which integrates \textsc{Spatial-AST} with the LLaMA-2 LLM, resulting in a model capable of answering complex reasoning questions about multiple sound sources situated within a 3-D environment.
\end{itemize}

\begin{table*}[ht]
    \centering
    \footnotesize
    \setlength\tabcolsep{1.5pt}
    \caption{\dataname~Overview: The first four types focus on perception, while the last emphasizes reasoning. DP: Distance Prediction; DoA: Direction-of-Arrival. Numbers (e.g., 139K, 15.9\%) indicate the QA sample count and their percentages in the dataset.}
    \label{tab: data_overview}
    \begin{tabular}{lcl}
    \toprule
       \textbf{Type} & \textbf{\#Sources} & \textbf{Example} \\
    \midrule
        \multirow{2}{*}{\shortstack[l]{\textbf{A:} Detection \\ (139K, 15.9\%)}} & \multirow{2}{*}{1} & \textbf{Q:} Identify the sound events in the audio clip. / \textbf{A:} baby laughter; laughter; speech \\ 
         & & \textbf{Q:} What are the distinct sounds present in this audio clip? / \textbf{A:} heart sounds, heartbeat \\
    \midrule
        \multirow{2}{*}{\shortstack[l]{\textbf{B:} DoA \& DP \\ (139K, 15.9\%)}} & \multirow{2}{*}{1}  & \textbf{Q:} How would you describe the location of this audio clip? / \textbf{A:} right, front, below; 2.5m \\
        & & \textbf{Q:} At what distance and in which direction, is the music's sound originating? / \textbf{A:} left, behind, below; 5m \\ 
    \midrule 
        \multirow{4}{*}{\shortstack[l]{\textbf{C:} Detection \\ (118K, 13.5\%) }} & \multirow{4}{*}{2} & \textbf{Q:} Identify the sound events in the audio clip coming from the right, front, below, approximately \\
        & & \quad \ 3 meters away. / \textbf{A:} slosh; speech \\
      & & \textbf{Q:} What sound events can you detect in the audio recording emanating from the left, behind, above, \\
      & & \quad \ roughly 0.5 meters away? / \textbf{A:} music; musical instrument; steelpan \\
    \midrule 
        \multirow{3}{*}{\shortstack[l]{\textbf{D:} DoA \& DP \\ (118K, 13.5\%)}} & \multirow{3}{*}{2} & \textbf{Q:} In which direction and how far away is the source of the heart sounds, heartbeat's sound? \\
        & & \textbf{A:} left, behind, below; 1m \\
        & & \textbf{Q:} Where is the sound of the music coming from? / \textbf{A:} left, behind, below; 3m \\
    \midrule 
        \multirow{8}{*}{\shortstack[l]{\textbf{E:} Reasoning \\ (358K, 41.2\%) }} & \multirow{8}{*}{2} &  \textbf{Q:} When measuring the direct line distance, is the sound produced by wheeze closer to you than \\ 
        & & \quad \ the sound from bird flight, flapping wings? / \textbf{A:} No \\
        & & \textbf{Q:} Is the source of both explosion and speech's sounds from your left side? / \textbf{A:} Yes \\
        & & \textbf{Q:} Is the sound from the music on the front and the sound from the rattle on the behind? / \textbf{A:} No \\
        & & \textbf{Q:} Does the sound of electric shaver, electric razor appear on the behind of the sound of waterfall? / \textbf{A:} Yes \\
        & & \textbf{Q:} Can you estimate the distance from the sound of the speech to the sound of the dog? / \textbf{A:} 1.5m \\
        & & \textbf{Q:} What is the sound on the above side of the sound of the vibration? / \textbf{A:} croak; frog \\
        & & \textbf{Q:} Could you determine whether the singing's sound is to the left or right of the steam's sound? / \textbf{A:} left \\
    \bottomrule
    \end{tabular}
\end{table*}

\section{Related Work}
\subsection{Spatial Audio}\label{sec: spatial_audio}
Spatial audio is a family of techniques used to create the illusion of sound sources existing within a 3-D space around the listener. Encapsulating traditional stereo and surround sound systems to more recent methods such as ambisonic audio, spatial audio is used in applications ranging from virtual reality~\cite{mystakidis2022metaverse} to advanced theater systems. In the realm of artificial intelligence and machine learning, spatial audio presents unique challenges for intelligent agents situated in 3-D physical spaces in accurately localizing and interpreting sound sources~\cite{evers2020locata, guizzo2022l3das22}. To address these challenges, researchers have developed acoustic simulation techniques~\cite{Scheibler_2018, chen2022soundspaces} and algorithms that leverage spatial audio~\cite{yang2022srp, wang2023fnssl}. Existing spatial audio datasets, such as YouTube-360~\cite{morgado2020learning}, YouTube-ASMR~\cite{yang2020telling}, Pano-AVQA~\cite{Yun2021PanoAVQA}, and STARSS23~\cite{Shimada2023starss23_arxiv}, offer varying levels of spatial audio information. However, they often suffer from inconsistent quality, lack crucial ground truth labels like sound source distance and direction, and are limited in scope, hindering the development of spatial audio-based machine learning models.

\subsection{Sound Event Localization and Detection}
The Sound Event Localization and Detection (SELD) task~\cite{adavanne2018sound} merges sound source localization with sound event detection (SED), a task which the DCASE community\footnote{\url{https://dcase.community/community_info}} later incorporated into their challenges as Task 3. Subsequent works~\cite{Du2020_task3_report,Zhang_UCAS_task3_report,Du_NERCSLIP_task3_report} adapted architectures like GRU~\cite{cho2014properties}, TDNNF~\cite{povey2018semi}, and Conformer~\cite{gulati2020conformer} for SELD tasks, achieving strong performance. However, these models have primarily focused on shallow spatial audio perception, and the potential of leveraging large language models to enable spatial reasoning remains unexplored in prior work, which is the main focus of our work.

\subsection{Multimodal Large Language Models}
Recent models~\cite{openai2023gpt4v,li2023blip2,liu2023visual,peng2023kosmos2} have equipped LLMs with the ability to reason about and create images and videos. For the audio modality, AudioGPT~\cite{huang2023audiogpt} integrates ChatGPT as a versatile interface for a wide array of audio and speech applications. Pengi~\cite{deshmukh2023pengi} proposes the use of transfer learning to frame all audio tasks as text-generation challenges, utilizing both audio and text inputs to produce results without additional fine-tuning. LTU~\cite{gong2023listen} proposes a monaural audio QA dataset, and combines an Audio Spectrogram Transformer (AST)~\cite{gong21b_interspeech} feature extractor with a LoRA-adapted~\cite{hu2021lora} LLaMA~\cite{touvron2023llama1} LLM to train the model to reason and answer questions about the sounds in a clip. Crucially, LTU differs from our work in that LTU is trained only on monaural sound data, whereas we consider multi-channel spatial sound rendered by a reverberant, 3-D environment. SALMONN~\cite{tang2023salmonn} uses OpenAI’s Whisper model~\cite{radford2023robust} and BEATs~\cite{chen2022beats} as dual auditory encoders to extract speech and audio representations, and then concatenates them to LLMs 
to enable the generation of responses. Qwen-audio~\cite{chu2023qwen} focuses solely on training the Whisper-based audio encoder using large-scale datasets, achieving proficiency in universal audio understanding across a diverse range of tasks and audio types. However, despite their impressive performance in the audio domain, none of these models have the capability to perceive and reasoning about spatial audio that is situated in diverse, reverberant, and complex 3-D environments.

\section{\dataname}
The training of \modelname~requires a spatial audio-based question answering dataset. Current datasets like SpatialVLM~\cite{chen2024spatialvlm} and Pano-AVQA~\cite{Yun2021PanoAVQA} are focused on vision and lack spatial audio information (or audio altogether). To fill this void, we introduce \dataname, the first spatial audio-based question answering dataset, which is designed to emphasize the complexities of the spatial audio, free from visual influences, and caters to the unique demands of spatial audio perception and reasoning.

\subsection{Spatial Audio Generation}


Collecting and annotating spatial audio in the real world is a time consuming task that is made more complex by environmental acoustic variability and limitations in recording equipment. In order to efficiently create a dataset with a large amount of variability across environments and types of sound sources that is also rich in ground-truth metadata about these sources, we have adopted a simulation-based approach for generating spatial audio data for the dataset.

\textbf{Spatial Audio Simulator.} We use the state-of-the-art audio simulator, SoundSpaces 2.0~\cite{chen2022soundspaces}. This platform performs on-the-fly geometry-based sound rendering, enabling realistic acoustic reverberation with arbitrary source-receiver locations. Many simulation parameters can be easily modified, such as the material properties of a room's walls and the objects within the room, as well as the geometry of the receiver's microphone array. This enables us to curate a diverse and highly realistic dataset that also offers easy access to the ground-truth sound generation parameters, such as the spatial location and orientation of each source within the environment. Specifically, we leverage Matterport3D~\cite{Matterport3D} for our environmental meshes, which includes highly detailed mesh renderings of 90 complete buildings, each featuring an average of 24.5 rooms across 2.61 floors with an average floorspace of 517.34 $m^2$. 
For a given arbitrary source location $s$, a monaural sound source $A^s$, receiver location $r$, and the receiver's heading direction $\theta$ in a specific mesh environment, the audio signal $A^r$ received by a microphone is given by the convolution of the room impulse response with the sound source, as shown in the following equation:
\begin{equation}
    A^r_m(t) = R_m(t, s, r, \theta) * A^s(t)
\end{equation}
where $t\in[1, T]$ represents the time sample index, $m \in [1, 2, ..., n]$ represent the microphone index and $R_m(t, s, r, \theta)$ is the impulse response between the source and receiver. 

To minimize perceptual dissonance in auditory localization when visual cues are absent, we ensure that both the sound source and the receiver are located within the same room. We position the receiver at coordinates that allow for an upright stance. As for the sound sources, they are placed at random locations throughout the room. In total, 21,131 reverberations are generated, averaging approximately 9.58 reverberations per room.

\textbf{Sound Sources.} Previous spatial audio datasets~\cite{yang2020telling, Shimada2023starss23_arxiv} often encompass a limited range of audio events, typically restricted to categories like music, speech, and domestic sounds. To broaden our scope and better represent real-world acoustic scenarios, we sample from  AudioSet~\cite{Audioset} to specify our monaural sound source $A^s$. AudioSet consists of roughly 2 million 10-second in-the-wild YouTube clips used for audio classification, with each clip weakly annotated with 527 types of audio events, often featuring multiple overlapping events within a single clip. However, certain categories among these 527 types are discernible only through visual cues (e.g., ``Single-lens reflex camera" vs. ``Camera"). To enhance the dataset's reliability, we exclude labels that require visual information by manual inspection. Additionally, to mitigate the impact of poor-quality training samples, we remove labels with less than 50\% quality.\footnote{For example, the estimated quality of the ``clatter" label is 20\%. Source: https://research.google.com/audioset/dataset/clatter.html.} Ultimately, given that the input audio undergoes convolution with reverberation, we further exclude most noise-related labels, such as ``Noise", ``Echo" and ``Outside, urban or manmade". This curation results in a set of 355 audio event labels that are appropriate for our setting, identifiable solely through audio cues. After filtering by these categories, we are left with 1,861,750 clips in the AudioSet-2M split, while our filtered AudioSet-20K split contains 18,373 clips, and our evaluation set consists of 17,148 clips. Lastly, to counteract any volume variability in the source audio, we applied loudness normalization across all clips.



\subsection{Question-Answer Pair Generation}
We curated \dataname~to include diverse question-answer pairs, all centered around the challenges of spatial audio perception and reasoning. Similar to LTU~\cite{gong2023listen}, each sample in the dataset is structured as a tuple of (audio, question, answer), where the audio and question serve as inputs to the model, and the answer acts as the target label. Table~\ref{tab: data_overview} illustrates the array of tasks in \dataname, ranging from basic perception to complex spatial reasoning. To enhance variety, the questions in these tasks are paraphrased using GPT-4, ensuring a broad spectrum of queries. The answers, on the other hand, are generated uniformly through a systematic rule-based approach, maintaining consistency across the dataset. We enumerate each question type in the following paragraphs.
\textbf{Sound Event Detection (Type A \& C).} For this task, the question is ``What sound events can you detect in the audio recording?" and its paraphrases. The answer may consist of a sorted list of sound classes present in the audio clip, arranged in alphabetical order. Question type A utilizes audio with 1 sound source, while C uses 2.

\textbf{Direction and Distance Estimation (Type B \& D).} This task aims to identify the direction and distance of a sound source. For establishing ground truth, the three-dimensional space is segmented into eight regions. Each region is defined by a tuple representing the directional axis (left/right, front/behind, above/below) with respect to the receiver. Additionally, distance is quantified in increments of 0.5 meters, spanning a range from 0 to 10 meters. A typical question might be, ``Where is the sound of the music coming from?" accompanied by GPT-generated paraphrased versions. The format for answers is constrained to expressions like ``left, front, above; 2.5m". Question type B utilizes audio with 1 sound source, while D uses 2.


\textbf{Spatial Reasoning (Type E).} This task differs from the previous perception-focused tasks by introducing scenarios where two sound events occur concurrently, originating from distinct distances and directions. When generating these scenarios, we ensure that each mix contains two sound sources with entirely different sound events and orientations. The challenge here is to discern the spatial differences between these overlapping sounds. For example, a question might be, ``What is the sound on the \texttt{left} side of the sound of the \textit{dog barking}?" This task demands both perception and complex reasoning. The model must implicitly separate the sound sources based on their unique classes, spatially localize each source, and then analyze the relationship between the sources in the context of the question. However, the encoder \textsc{Spatial-AST} was not been exposed to such mixed data types during its pre-training stage, implying a lack of audio separation ability. This task is thus designed to test the model's in-context learning capabilities and effectively utilize the latent representations from the encoder to address this complex multi-source audio reasoning challenge. To make evaluation straightforward, we have chosen a binary (Yes/No) response format for these questions. By adopting binary answers, we can more effectively gauge the model's reasoning performance and maintain consistent assessment standards across different question types.


\begin{figure*}[!htb]
    \centering
    \includegraphics[trim={0 0 25 0},clip, scale=0.75]{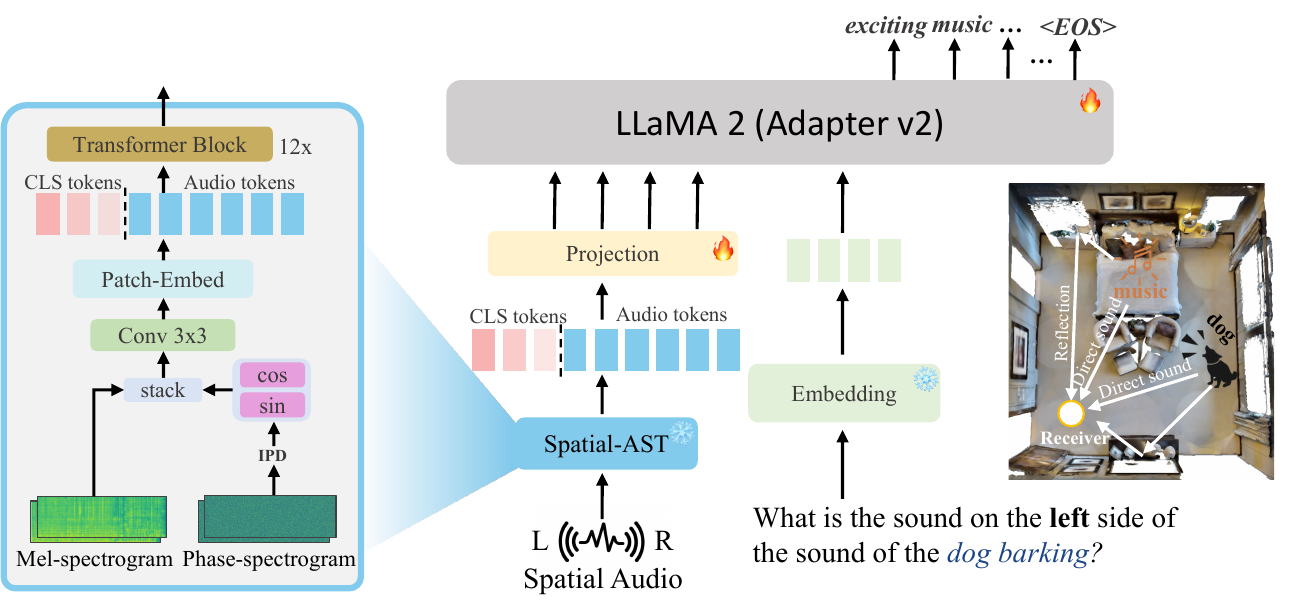}
    \caption{Illustration of the \modelname~model structure. The rightmost part shows the binaural audio generation process.}
    \label{fig:pipeline}
\end{figure*}
\section{Method}
Figure~\ref{fig:pipeline} illustrates our overall architecture for the \modelname~model. We will first introduce \textsc{Spatial-AST}, our novel spatial audio encoder that acts as a front end for \modelname. We then discuss how it is integrated with the LLaMA-2 LLM in order to build spatial reasoning ability on top of \textsc{Spatial-AST}'s auditory perception ability. 

\subsection{Spatial Audio Encoder: \textsc{Spatial-AST}}\label{section: encoder}
We propose a novel architecture \textsc{Spatial-AST} to capture spatial audio information. The model is shown on the left-hand side of Figure~\ref{fig:pipeline}.

\paragraph{Front-End.} 
Our encoder integrates both Mel-Spectrogram and Interaural Phase Difference (IPD). As detailed in the \textsc{Spatial-AST} section of Figure~\ref{fig:pipeline}, we initially transform the time-domain signal $x(n)$ into the frequency domain $X(t, f)$ using the short-time Fourier transform (STFT), as represented by Equation~\ref{eq: stft}:
\begin{equation}\label{eq: stft}
X_i(t, f) = \sum_{n=0}^{N-1} x_i(n) w(n-t) e^{-j\frac{2 \pi f n}{N}},\  i \in \{1, 2\}
\end{equation}
Here $w(n)$ denotes an N-point window function, and $i$ indicates the left or right channel in the binaural recording. 

Subsequently, we compute both the Mel-Spectrogram and the Interaural Phase Difference (IPD) based on these frequency domain signals $X_i(t, f)$. The Mel-Spectrogram for each channel is calculated as per Equation~\ref{eq: mel}, and the IPD is derived from the phase spectrograms of the left and right channels, as indicated in Equation~\ref{eq: ipd}:
\begin{equation}\label{eq: mel}
S_i(t, m) = log\left(|X_i(t, f)|^2 \times \text{melW} \right)
\end{equation}
where melW is a $M$-bin Mel filter bank.
\begin{equation}\label{eq: ipd}
IPD(t, f)=\angle \frac{X_2(t, f)}{X_1(t, f)}
\end{equation}

To address the numerical instabilities due to phase wraparound in IPD, we apply cosine and sine transformations, which are then weighted by melW to align with the dimension of the Mel-spectrogram. The final front-end output, $\mathcal{Z}$, is a concatenation of these processed components. It includes both the Mel-Spectrograms of the left and right channels, as well as the cosine and sine transformations of the IPD, all combined as outlined in Equation~\ref{eq: concate}:

\begin{equation}\label{eq: concate}
\begin{split}
\mathcal{Z} = [&S_1; S_2; \cos(\text{IPD}) \times \text{melW}; \\
&\sin(\text{IPD}) \times \text{melW}], \quad \mathcal{Z} \in \mathbb{R}^{4 \times T \times M}
\end{split}
\end{equation}

\textbf{Backbone.} The initial input $\mathcal{Z}$ is first passed to a 3x3 2D convolution followed by a batch normalization~\cite{ioffe2015batch} layer and a GELU~\cite{hendrycks2016gaussian} layer to fuse the inter-channel information. Subsequently, we employ a Patch-Embed CNN with a kernel size of $(16 \times 16)$ and corresponding strides in both time and frequency dimensions, ensuring non-overlapping segmentation of the features into distinct patch tokens. To extract spatial cues, we concatenate three [CLS] tokens at the beginning of the audio tokens, each specifically designated for extracting information about the audio's category, distance, and direction respectively.
All tokens are then fed into 12-layer Transformer encoder blocks. Finally, we process the three [CLS] tokens using three separate linear layers to get predictions for their corresponding tasks.

\textbf{Pre-Training Objective.} The spatial encoder is pre-trained to handle three tasks, namely sound event detection, distance prediction, and directional prediction, with the latter being further divided into azimuth and elevation angle predictions. We employ cross-entropy loss for all three tasks. To discretize prediction targets for distance and directional prediction, we categorize distance into intervals of 0.5 meters and angle into intervals of 1 degree for both azimuth and elevation.


The final pre-training objective of the spatial encoder is:

\begin{equation}\label{eq:encoder_loss}
    \mathcal{L} = \lambda_1 \mathcal{L}_{cls} + \lambda_2 \mathcal{L}_{dis} + \lambda_3 \mathcal{L}_{doa}
\end{equation}
where $\mathcal{L}_{cls}$, $\mathcal{L}_{dis}$, $\mathcal{L}_{doa}$ represent the losses for detection, distance, and direction respectively, while $\lambda_1$, $\lambda_2$, $\lambda_3$ are hyperparameters.

\subsection{ \modelname:~A Model for Reasoning about Spatial Sounds}
On its own, \textsc{Spatial-AST} is capable of predicting the type, direction, and distance for an input spatial sound. To extend this ability to encompass spatial reasoning about multiple sound sources present within an environment, we fuse \textsc{Spatial-AST} to the LLaMA-2 7B LLM~\cite{touvron2023llama2}. Input spatial audio is first processed by the \textsc{Spatial-AST} encoder, and then a projection module is used to map its set of output tokens into LLaMA-2's input text embedding space. For efficient fine-tuning, we use the LLaMA-adapter v2~\cite{gao2023llama}.

The fine-tuning objective of \modelname~is to predict the answer text conditioned on its paired question text and the corresponding audio input. The model seeks to maximize the probability of predicting the next answer token, applying cross-entropy loss for each token position $1 \leq \tau \leq T$ in the sequence, which can be mathematically represented as:
\begin{equation}
 \mathcal{P}_\theta\left(y_{\tau} \mid y_{\tau-1}, \cdots, y_1; a_l, a_{l-1}, \cdots, a_1\right)
\end{equation}
where $a_l$ is the \textit{l}-th token of audio embedding, and $y_{\tau}$ denotes the \textit{$\tau$}-th token of text output.
\begin{table}
    \centering
    \caption{The perception-to-reasoning curriculum.} \label{tab: llm_curriculum} 
    \begin{tabular}{cccc}
\toprule
    Stage & Question Type & \# Tr. Samples & Percentages\\ 
\midrule
    \uppercase\expandafter{\romannumeral1} & A, B & 278K & 31.8\%\\
    \uppercase\expandafter{\romannumeral2} & A, B, C, D & 514K & 58.8\%\\
    \uppercase\expandafter{\romannumeral3} & A, B, C, D, E & 872K & 100\%\\
\bottomrule
\end{tabular}
\end{table}

For the training of \modelname, we devised a perception-to-reasoning curriculum, as detailed in Table~\ref{tab: llm_curriculum}. The process begins with a ``warm-up" phase focusing on single-source perception tasks (question types A and B), enabling the model to adapt to audio modality inputs and infer basic properties of a single audio source including its class, distance, and direction. In the second stage, we introduce multi-source scenarios (question types C and D), where the model is taught to implicitly perform sound source separation by answering questions about specific sources within mixed audio. The final stage unlocks the full capabilities of \modelname, integrating reasoning tasks (question type E), such as interpreting spatial relationships and distances between multiple sound sources. The importance of this curriculum is evidenced in Table~\ref{tab: bat_curriculum} in Appendix~\ref{appendix: bat}.

\section{Experiments}


\subsection{Implementation Details}
\textbf{Input Audio Processing.}
The 10-second audio sources are first loudness normalized by scaling them so that each clip has the same total sound energy. After convolution with the room impulse response from SoundSpaces, we trim or pad these clips to 10 seconds to align with AudioMAE~\cite{huang2022masked}. The resulting waveforms are binaural with 2 channels at a 32kHz sampling rate. We use a window size of 1024, a hop size of 320, and 128 mel-bins to compute the Short-Time Fourier Transforms (STFTs) and mel-spectrograms. As a result, for a 10-second recording from AudioSet, the concatenated Mel-spectrogram and IPD feature dimension is (4, 1024, 128). 

\textbf{Spatial Audio Encoder.}
We implement a patch masking ratio of 0.25 in both time and frequency during training, and the unmasked tokens are then concatenated with three learnable [CLS] tokens. We initialize the weights of the transformer blocks using the official pretrained AudioMAE~\cite{huang2022masked} checkpoint. The encoder is trained on 8 RTX 3090 GPUs, with each epoch taking approximately 10 minutes. Recognizing the greater challenge posed by sound event detection compared to distance and directional prediction, we train our encoder in two stages. In the first stage, we focus exclusively on detection loss, setting the weights $\lambda_2$ and $\lambda_3$ to zero as per Equation~\ref{eq:encoder_loss}. In the second stage, we reintroduce the losses for distance and direction, adjusting the loss weights to $\lambda_1=1250, \lambda_2=1$, and $\lambda_3=2$ to broaden the capabilities of the model. This two-stage pre-training approach is crucial for achieving a balanced performance across different tasks. The effects of varying these weights on model performance for various tasks are detailed in Appendix~\ref{appendix: encoder}.

\textbf{LLM.} 
Diverging from the original two-stage learnable parameters design of LLaMA-adapter v2, We concurrently train parameters including zero-initialized attention, projection, norm, bias, and scale to simplify the training process. 
The training is completed on 8 V100 GPUs. For generation, we employ a greedy decoding and set the temperature to 0.1 and nucleus sampling~\cite{holtzman2019curious} top p to 0.75.

\begin{table*}[!htb]
    \centering
    \setlength\tabcolsep{3pt}
    \caption{This table presents a comparative analysis of various models using both monaural and binaural data, focusing on key performance metrics: mean Average Precision (mAP, $\uparrow$), Error Rate at 20° (ER\textsubscript{20°}, $\downarrow$), Mean Angular Error (MAE, $\downarrow$), and Distance Error Rate (DER, $\downarrow$). Results for AudioMAE models were derived through inference using their official checkpoints, without further training. Models that utilize anechoic audio data (no reverberation) are \textcolor{gray!50}{grayed-out} for clarity.}
    \label{tab: encoder}
    \begin{tabular}{@{}cccccccc@{}}
    \toprule
         & & \multicolumn{2}{c}{Data} & \multicolumn{4}{c}{Performance Metrics} \\
         \cmidrule(lr){3-4} \cmidrule(lr){5-8}
         & &  Audio Type & Input Features & mAP ($\uparrow$) & ER\textsubscript{20°} ($\downarrow$) & MAE ($\downarrow$) & DER ($\downarrow$) \\
        \midrule 

        \multicolumn{2}{c}{\textcolor{gray!65}{AudioMAE}~\cite{huang2022masked}} & \textcolor{gray!65}{Anechoic} & \textcolor{gray!65}{Mel-spectrograms} & \textcolor{gray!65}{53.56} & \textcolor{gray!65}{-} & \textcolor{gray!65}{-} & \textcolor{gray!65}{-} \\ \midrule
        
        \multicolumn{2}{c}{AudioMAE~\cite{huang2022masked}} & \multirow{4}{*}{Monaural} & \multirow{4}{*}{Mel-spectrograms} & 47.18 & - & - & - \\ 
        
        \multirow{3}{*}{\begin{tabular}[c]{@{}c@{}}\quad \textsc{Spatial-AST} (ours)\end{tabular}} & Joint-Training & & & 51.39 & 95.85 & 88.52 & \textbf{26.87} \\ 
        
        & Stage-1 & & & 52.26 & - & - & - \\
        & Stage-2 & & & 50.69 & 95.35 & 89.42 & 28.25 \\ \midrule

        \multicolumn{2}{c}{SELDnet~\cite{adavanne2018sound}} & \multirow{3}{*}{Binaural} & Mel-spectrograms, IPD & 42.66 & 25.19 & 19.21 & 38.46 \\ 
        \multirow{1}{*}{\begin{tabular}[c]{@{}c@{}}\quad \quad \textsc{Spatial-AST} (ours)\end{tabular}} & & & Mel-spectrograms & 49.94 & 25.50 & 18.57 & 34.85 \\
        \multirow{1}{*}{\begin{tabular}[c]{@{}c@{}}\quad \quad \textsc{Spatial-AST} (ours)\end{tabular}} & \multirow{-2}{*}{Joint-Training} & & Mel-spectrograms, IPD & 50.57 & 24.81 & 18.16 & 35.29 \\ \midrule        
        \multirow{2}{*}{\begin{tabular}[c]{@{}c@{}}\quad \quad \textsc{Spatial-AST} (ours)\end{tabular}} & Stage-1 & \multirow{4}{*}{Binaural} & \multirow{2}{*}{Mel-spectrograms} & 51.70 & - & - & - \\
        & Stage-2 & & & 50.86 & 30.94 & 21.66 & 28.60 \\ \cmidrule(lr){1-2} \cmidrule(lr){4-4}
        
        \multirow{2}{*}{\begin{tabular}[c]{@{}c@{}}\quad \quad \textsc{Spatial-AST} (ours)\end{tabular}} & Stage-1 & & \multirow{2}{*}{Mel-spectrograms, IPD} & \textbf{52.26} & - & - & - \\
        & Stage-2 & & & 50.03 & \textbf{23.89} & \textbf{17.94} & 32.54 \\ 

        \bottomrule
    \end{tabular}
\end{table*}

\textbf{Baseline.}
For our encoder baseline comparisons, we select established models such as AudioMAE~\cite{huang2022masked} and SELDnet~\cite{adavanne2018sound}. To ensure a fair comparison, SELDnet's feature extraction network is augmented with a 12-layer transformer block, aligning the model parameters of the compared architectures to approximately 90M. Additionally, for \modelname, we use a version with monaural input for comparison, which has an architecture similar to LTU~\cite{gong2023listen}, serving as a baseline.

\textbf{Evaluation Metrics.}
In the evaluation of our spatial audio encoder, outlined in Table~\ref{tab: encoder}, we use the mean Average Precision (mAP) for sound event detection; Error Rate at 20° (ER\textsubscript{20°}), as referenced in~\cite{mesaros2019joint}, and Mean Angular Error (MAE) for Direction of Arrival (DoA) accuracy, evaluating whether predictions deviate more than 20° from the reference point and quantifying the average angular deviation between predictions and ground truth; and Distance Error Rate (DER) for measuring the accuracy of distance predictions within 0.5 meters of the actual location. For the evaluation of \modelname, as shown in Table~\ref{tab: LLM_performance}, we apply similar metrics, including mAP and DER. For the DoA task, where the three-dimensional space is segmented into eight distinct sections, we use accuracy (Acc) as the performance metric. Additionally, for reasoning tasks in question type E, Binary Accuracy (BA) serves as the evaluation metric. 

\subsection{Performance of \textsc{Spatial-AST} on SELD Tasks}


Table~\ref{tab: encoder} presents the experimental results for our proposed \textsc{Spatial-AST} audio encoder. We compare \textsc{Spatial-AST} against current state-of-the-art models such as AudioMAE~\cite{huang2022masked} and SELDnet~\cite{adavanne2018sound}. We also compare spatial cue extraction between monaural and binaural data. Additionally, we explore the impact of two pre-training methodologies, including joint-training and two-stage approaches, and highlight the importance of IPD.

\begin{table*}[!htp]
    \centering
    \caption{Performance evaluation on \dataname. The metrics include mAP for detection accuracy, represented by two columns corresponding to question types \textcolor{gray!65}{A} and C, respectively; Accuracy (Acc) for assessing the accuracy of identifying directions and Distance Error Rate (DER) for distance prediction, with two columns each for question types \textcolor{gray!65}{B} and D; and Binary Accuracy (BA) for the accuracy of binary (Yes/No) responses in reasoning (question type E). Input types include monaural audio (M), binaural audio (B), prompt (P), or ground truth sound source parameters (GT). The ``Random" and ``Oracle" rows serve as baseline and ideal performance benchmarks.} \label{tab: LLM_performance} 
    \renewcommand{\arraystretch}{1.2} 
    \begin{threeparttable}
    \begin{tabular}{@{}ccccccc|c@{}}
\toprule
 \multirow{3}{*}{Model} & \multirow{3}{*}{Input} & \multicolumn{3}{c}{Perception (Type \textcolor{gray!65}{A}C\textcolor{gray!65}{B}D)\tnote{$\dag$}} & \multicolumn{3}{c}{Reasoning (Type E)} \\ \cmidrule(lr){3-5} \cmidrule(lr){6-8}
  & & Detection (mAP~$\uparrow$)\tnote{*} & DoA (Acc~$\uparrow$) & DP (DER~$\downarrow$) & Direction & Distance & Avg. \\
     
\midrule 
Random & P & \textcolor{gray!65}{0.61} \textbar{} 0.59 & \textcolor{gray!65}{12.57} \textbar{} 12.41 & \textcolor{gray!65}{67.33} \textbar{} 67.46 & 50.00 & 50.00 & 50.00 \\

\midrule
Monaural \modelname~(ours)\tnote{1} & M + P & \textcolor{gray!65}{24.15} \textbar{} 6.42 & \textcolor{gray!65}{14.31} \textbar{} 11.93 & \textcolor{gray!65}{34.17} \textbar{} 56.26 & 57.69 & 51.36 & 54.53 \\

\midrule
One-stage \modelname\tnote{1} & \multirow{2}{*}{B + P} & \textcolor{gray!65}{26.23} \textbar{} 9.06 & \textcolor{gray!65}{\textbf{76.49}} \textbar{} 37.09 & \textcolor{gray!65}{29.28} \textbar{} 51.62 & 65.12 & 81.98 & 73.55 \\

One-stage \modelname\tnote{1, 2} & & \textcolor{gray!65}{25.50} \textbar{} 5.67 & \textcolor{gray!65}{72.74} \textbar{} 35.63 & \textcolor{gray!65}{31.51} \textbar{} 73.68 & 59.59 & 49.76 & 54.68 \\
   
\hdashline

\multirow{2}{*}{\modelname} & P & \textcolor{gray!65}{0.69} \textbar{} 0.65 & \textcolor{gray!65}{12.88} \textbar{} 13.14 & \textcolor{gray!65}{51.65} \textbar{} 71.73 & 58.28 & 50.67 & 54.48 \\

& B + P & \textcolor{gray!65}{\textbf{26.34}} \textbar{} \textbf{9.89} & \textcolor{gray!65}{75.54} \textbar{} \textbf{37.65} & \textcolor{gray!65}{\textbf{29.16}} \textbar{} \textbf{47.90} & \textbf{69.77} & \textbf{84.01} & \textbf{76.89} \\

\midrule
\rowcolor{gray!25} Oracle\tnote{1} & GT + P & \textcolor{gray!65}{94.16} \textbar{} 94.26 & \textcolor{gray!65}{99.20} \textbar{} 99.81 & \textcolor{gray!65}{0.00} \textbar{} 0.00 & 98.21 & 100.00 & 99.10 \\ 
\bottomrule
\end{tabular}
\begin{tablenotes}
\item[$\dag$] We grey out the columns corresponding to question types \textcolor{gray!65}{A} and \textcolor{gray!65}{B}, where the model receives a single sound source.
\item[*]\footnotesize{Like LTU~\cite{gong2023listen}, we use a text encoder (text-embedding-3-small) to encode \modelname's output and the labels from the evaluation dataset, then compute the cosine similarity between these text embeddings for evaluation. mAP potentially underestimates \modelname's performance.}
\item[1]\footnotesize{Only train stage \uppercase\expandafter{\romannumeral3}.} \qquad \textsuperscript{2} We removed question types C and D from \dataname.
\end{tablenotes}
\end{threeparttable}
\end{table*}

\textbf{Monaural vs. Binaural.} To support monaural input, we modified the 3x3 convolution module, setting its input channel to 1, and only feed a monaural mel spectrogram as input. Despite binaural data typically offering richer spatial information, our encoder demonstrates robust performance with monaural inputs, especially in sound event detection (SED) and distance prediction (DP) tasks. The encoder achieves a mAP of 51.39 for monaural input, surpassing both its binaural counterpart, which scores 49.94, and AudioMAE, which achieves 47.18 in mAP. Additionally, our encoder with monaural input even attains the best Distance Error Rate (DER) compared to models using binaural input. However, as monaural audio cannot capture spatial information, it shows low performance in the direction-of-arrival (DoA) task, with an ER\textsubscript{20°} of 95.82 and an MAE of 88.52. The performance of monaural data suggests it is sufficiently effective for SED and DP tasks. The effectiveness of monaural input can be attributed to the fact that binaural audio does not necessarily provide additional benefits for sound category perception. The results from the single-task setting detailed in Appendix~\ref{appendix: single-task} corroborate this conclusion.



\textbf{Joint-training vs. Two-stage training.} Joint-training emerges as the more effective approach for monaural data, outperforming two-stage training in both detection and distance prediction tasks, achieving a mAP of 51.39 and DER of 26.87. However, for binaural data, two-stage training consistently demonstrates improvements in its second stage, regardless of whether IPD is used. The enhancement is particularly significant with the usage of IPD, achieving the lowest MAE of 23.89 and DER of 17.94, demonstrating the importance of including multi-microphone phase information for spatial audio perception. This also underscores the importance of employing a two-stage training approach for enhanced performance in binaural settings.

\textbf{Model Architecture.} When compared to AudioMAE and SELDnet, our encoder exhibits enhanced performance in reverberant environments. As AudioMAE is tailored for monaural input, we focus our comparison on monaural scenarios. In these contexts, our proposed model surpasses both joint-training and two-stage training methods. This achievement is especially significant given that our model has been finetuned for much fewer training hours. For simplicity, we compare our model with SELDnet in joint-training, where it surpasses SELDnet on all metrics.

\subsection{Performance of \modelname~on \dataname}

Table~\ref{tab: LLM_performance} shows the performance of various configurations of our \modelname~model on \dataname, broken down across the different question types. \modelname~performs strongly across all question types, outperforming the random baseline by a very large margin. We observe that the one-stage \modelname~generally lags behind the three-stage \modelname~in performance across various metrics. Intriguingly, we observed a significant drop in the model's reasoning ability when question types C and D, related to sound source separation, were removed from the training set. This is evident from the Binary Accuracy (BA) results, which approximate random chance in the reasoning tasks. Analysis of the one-stage \modelname's outputs reveals that the model struggles to differentiate between sound sources and their respective positions when confronted with reasoning questions. This finding highlights the importance of sound source separation tasks in training \modelname~to effectively handle complex reasoning, underscoring the necessity of a multi-stage curriculum that gradually introduces the model to increasingly sophisticated spatial audio perception and reasoning tasks.

The ``Monaural \modelname" row shows the performance of \modelname~trained with monaural audio. The architecture of monaural \modelname~is similar to that of LTU~\cite{gong2023listen}, so we use it as a baseline for comparison. The results reveal that monaural \modelname~faces significant challenges in reasoning tasks, achieving a Binary Accuracy (BA) of only 54.53\%. This suggests that monaural audio input alone cannot provide sufficient spatial information for complex reasoning challenges, highlighting the usefulness of \dataname~for learning spatial audio reasoning capabilities. More ablations can be found in Appendix~\ref{appendix: bat}.

\section{Limitations and Future Work}
We are confident that \modelname~will significantly contribute to the development of spatial audio perception and reasoning as well as multimodal LLM. However, as with any research tool, it is crucial to acknowledge certain limitations and assumptions inherent in our approach.

One of the primary challenges we encounter is extracting as much information as possible from spatial audio, with a specific focus on how to better utilize phase information. Currently, our model handles a maximum of two sound sources, with an emphasis primarily on audio. Looking ahead, there is potential to expand into multi-source scenarios and to integrate both audio and speech processing for a more holistic approach. Additionally, while our current framework is limited to binaural audio, exploring ambisonics could provide a more immersive and realistic spatial audio experience. Moreover, expanding \dataname~to be more open-ended would better align with human usage patterns and preferences, allowing for a wider range of queries and responses.

Another important consideration is the integration of additional modalities, such as visual information, to complement auditory cues. This multimodal approach might enhance the model's understanding and interpretation of complex environments. Lastly, the sim2real gap remains an aspect that requires further investigation. Observing how our model performs in real-world scenarios, as opposed to simulated environments, will be crucial in assessing its practical applicability and effectiveness.

\section{Conclusion}
In this work, we have presented \modelname, the first LLM capable of processing spatial audio input. To train and evaluate \modelname, we introduced \dataname, the first extensive spatial audio-based question answering dataset. We also proposed \textsc{Spatial-AST}, a novel spatial audio encoder capable of efficiently handling sound event detection, spatial localization, and distance estimation. \modelname~and \dataname~showcase the immense potential of LLMs for reasoning about spatial sound. \dataname~also enables rich future work, such as reasoning about the environment itself (materials, shape, layout), modeling moving sounds to for tracking problems, or incorporating the visual modality which SoundSpaces2.0 natively supports. 

\section*{Acknowledgements}
We thank Changan Chen, Linfeng Feng, and Jin Sob Kim for their constructive help. This work was supported in part by various grants and institutions. Zhisheng Zheng was supported by Zhiyuan College at Shanghai Jiao Tong University. Puyuan Peng and David Harwath's work was supported by the National Science Foundation under Grant No. 2238605. Xie Chen was partially supported by the National Natural Science Foundation of China (No. U23B2018, No. 62206171), the Shanghai Municipal Science and Technology Major Project under Grant 2021SHZDZX0102, and the International Cooperation Project of PCL.

\section*{Impact Statement}
Our research on \modelname, the first spatial audio-based large language model, alongside the \dataname~dataset, holds significant potential for broad impact across various fields:

\textbf{Advancements in Spatial Audio:} Our work addresses a key gap in spatial audio perception and reasoning. By developing an LLM specifically for spatial audio, we open up new possibilities for applications where spatial sound cues play a crucial role, such as virtual reality, gaming, and audio engineering. This can lead to more immersive and realistic experiences in these domains.

\textbf{Enhancements in Multimodal Large Language Models:} The integration of spatial audio into LLMs represents a significant step towards truly multimodal AI systems. Our model not only demonstrates the capability of LLMs in processing complex spatial audio information but also paves the way for future research in combining auditory and other sensory modalities, such as visual or tactile, to create even more sophisticated AI models.

\textbf{Empowering Embodied AI:} The ability to interpret and reason about spatial sounds can significantly enhance embodied AI systems, like robots or autonomous vehicles. These systems can utilize spatial audio cues for better navigation and interaction with their environment, leading to more effective and safer applications in real-world scenarios.



\newpage
\bibliography{icml2024}
\bibliographystyle{icml2024}

\newpage
\appendix
\onecolumn

\section{Why Do We Use Large Language Models?}
The use of Large Language Models (LLMs) in our spatial audio reasoning framework is necessary due to several critical reasons. First, LLMs enable complex questions to be posed using natural language, allowing for a more flexible and intuitive query format, such as ``Is the dog further away from you than the stereo?" or ``Is the dog to the left or to the right of the stereo?" This capability eliminates the need for a cascade of separate processing steps like source separation, classification, DoA, distance classifiers, and rule-based templates, which would significantly limit the types of questions that could be addressed. Additionally, LLMs facilitate the joint training of the spatial audio encoder and reasoning module in an end-to-end manner, reducing the risk of error propagation inherent in segmented processing approaches. Moreover, LLMs offer the ability to generalize across different questions with overlapping expressions, enhancing the model's adaptability and versatility in responding to a wide range of spatial audio-related queries. Last but not least, this task also intends to further expand the application boundaries of LLMs, demonstrating their versatility and effectiveness in domains beyond traditional text and image processing.



\section{Hyperparameters}
We present the specific training hyperparameter configurations for \textsc{Spatial-AST} \& \modelname~in Table~\ref{eq: hyperpara}.
\begin{table}[!htp]
\caption{Training hyperparameters of \textsc{Spatial-AST} \& \modelname}\label{eq: hyperpara}
\centering
\begin{tabular}{c|cc|ccc}
\toprule
\multirow{2}{*}{Configuration} & \multicolumn{2}{c|}{\textsc{Spatial-AST}} & \multicolumn{3}{c}{\modelname} \\
& Stage-1 & Stage-2 & Stage-\uppercase\expandafter{\romannumeral1} & Stage-\uppercase\expandafter{\romannumeral2} & Stage-\uppercase\expandafter{\romannumeral3} \\ 
\midrule
Sound Source & \multicolumn{2}{c|}{AudioSet-2M} & \multicolumn{3}{c}{AudioSet-20K} \\ 
Audio Normalization & \multicolumn{2}{c|}{\CHECK} & \multicolumn{3}{c}{\CHECK} \\ 
Augmentation & \multicolumn{2}{c|}{\CHECK} & \multicolumn{3}{c}{\CROSS} \\
Weighted sampling & \multicolumn{2}{c|}{\CHECK} & \multicolumn{3}{c}{\CROSS} \\
Optimizer & \multicolumn{5}{c}{AdamW~\cite{loshchilov2017decoupled}} \\ 
Optimizer momentum & \multicolumn{5}{c}{$\beta_1=0.9, \beta_2=0.95$} \\
Weight decay & \multicolumn{2}{c|}{0.0001} & \multicolumn{3}{c}{0.05} \\
Base learning rate & \multicolumn{2}{c|}{0.001} & \multicolumn{3}{c}{0.001} \\
Learning rate schedule & \multicolumn{5}{c}{half-cycle cosine decay~\cite{loshchilov2016sgdr}} \\
Warm-up epochs & 10 & 5 & \multicolumn{3}{c}{2} \\
Epoch partitioning factor & \multicolumn{2}{c|}{10} & \multicolumn{3}{c}{10} \\
Epochs & 60 & 40 & 2 & 2 & 3 \\
Batch size & \multicolumn{2}{c|}{64} & \multicolumn{3}{c}{2} \\
GPUs & \multicolumn{2}{c|}{8 RTX 3090} & \multicolumn{3}{c}{8 Tesla V100} \\

\bottomrule
\end{tabular}
\end{table}

\section{Room Acoustics Characteristics}
Table~\ref{tab: rt60} presents the average RT60 values across some common room types in the Matterport3D~\cite{Matterport3D} training settings. RT60 is a standard acoustic measurement that is defined as the time it takes for the sound pressure level to reduce by 60 dB~\cite{hak2012measuring}. This metric varies across different rooms due to factors such as room size, object arrangement, and construction materials, each influencing the room's acoustic characteristics. 

\begin{table}[h]
\caption{Average reverberation time (RT60) for different rooms.}
\centering
\begin{tabular}{c|c|c|c|c|c|c|c|c|c}
\toprule
 & Living Room & Bedroom & Closet & Bathroom & Toilet & Garage & Lounge & Kitchen & Dining Room\\
\midrule
RT60 & 0.2076 & 0.1917 & 0.1950 & 0.1906 & 0.1706 & 0.2190 & 0.2224 & 0.2061 & 0.2201\\
\bottomrule
\end{tabular}  \label{tab: rt60}
\end{table}

\section{Performance of \textsc{Spatial-AST} in a Single-Task Setting}\label{appendix: single-task}
We evaluate the performance of our encoder in single-task scenarios, specifically focusing on classification, distance prediction, and direction-of-arrival (DoA). As shown in Table~\ref{tab: single-task}, we analyze the impact of including or omitting Interaural Phase Difference (IPD), as well as comparing monaural versus binaural formats, on the performance of each task. Our findings reveal that IPD generally enhances performance across all tasks, indicating the value of spatial information. Moreover, we observe that monaural data provides sufficient information for tasks like sound event detection and distance prediction, achieving the best performance in Distance Prediction and comparable results to binaural data in Sound Event Detection. However, binaural data significantly improves performance in tasks involving direction perception, underscoring the encoder's potential to handle multiple tasks effectively and the importance of audio format in spatial audio perception.
\begin{table*}[!htb]
    \centering
    \caption{Performance of the encoder in single-task scenarios: classification, distance prediction, and Direction-of-Arrival (DoA). We compare the impact of including or excluding Interaural Phase Difference (IPD) on the performance of each task. Key evaluation metrics include mean Average Precision (mAP, $\uparrow$), Error Rate at 20° (ER\textsubscript{20°}, $\downarrow$), Mean Angular Error (MAE, $\downarrow$), and Distance Error Rate (DER, $\downarrow$).} 
    \begin{tabular}{@{}ccccccccc@{}}
    \toprule
         \multirow{2}{*}{Task} & \multicolumn{2}{c}{Data} & \multicolumn{4}{c}{Performance Metrics} \\ \cmidrule(lr){2-3} \cmidrule(lr){4-7}
          & Audio Format & Input Features & mAP (\%) $\uparrow$ & ER\textsubscript{20°} (\%) $\downarrow$ & MAE (°) $\downarrow$ & DER (\%) $\downarrow$ \\
        \midrule
        Detection & \multirow{3}{*}{Monaural} &  \multirow{3}{*}{Mel-spectrogram} & 52.13 & - & - & -\\
        Direction-of-Arrival & & & - & 96.80 & 85.47 & -\\ 
        Distance Prediction & & & - & - & - & \textbf{21.44} \\ \midrule
        
        \multirow{2}{*}{Detection} & \multirow{7}{*}{Binaural} &  Mel-spectrogram & 51.70 & - & - & - \\ 
         & & Mel-spectrogram, IPD & \textbf{52.26} & - & - & - \\ \cmidrule(lr){1-1} \cmidrule(lr){3-3}  
        
        \multirow{2}{*}{Direction-of-Arrival} & & Mel-spectrogram & - & 21.42 & 15.98 & - \\ 
         & & Mel-spectrogram, IPD & - & \textbf{20.11} & \textbf{15.36} & - \\ \cmidrule(lr){1-1} \cmidrule(lr){3-3}
         
        \multirow{2}{*}{Distance Prediction} & & Mel-spectrogram & - & - & - & 25.55\\ 
         & & Mel-spectrogram, IPD & - & - & - & 22.60\\ 
         \midrule
         Generalist &Binaural&Mel-spectrogram, IPD &50.03 &23.89 &17.94 &32.54\\
        
        \bottomrule
    \end{tabular}\label{tab: single-task}
\end{table*}

\section{Ablations of the Spatial Audio Encoder \textsc{Spatial-AST}}\label{appendix: encoder}
\begin{figure}[!htb]
    \centering
    \includegraphics[trim={40 10 30 0},clip, scale=0.37]{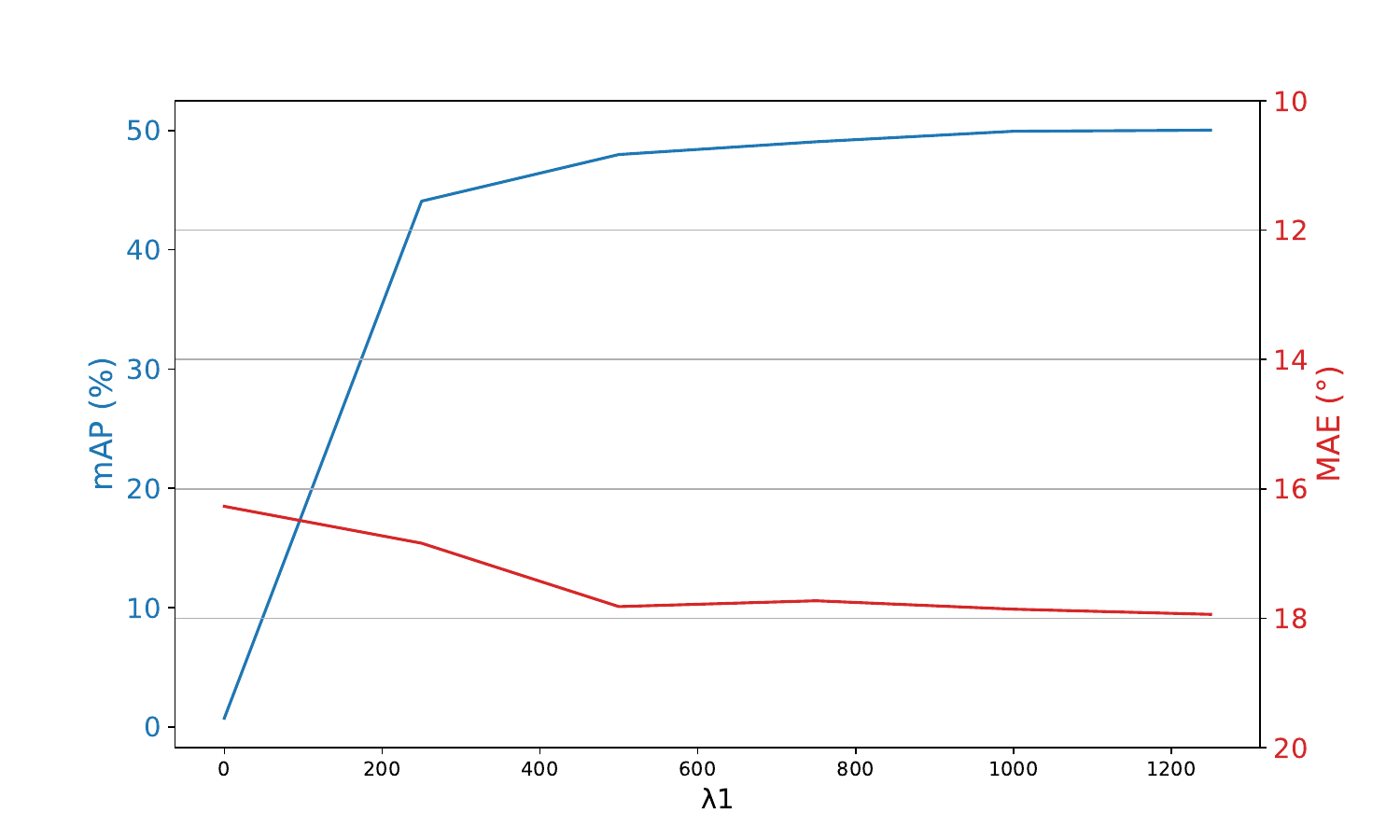}
    \caption{Classification (mAP) and Direction-of-Arrival (MAE) for different $\lambda_1$ values.}\label{fig: lambda_1}
\end{figure}

\textbf{Loss Weights.}
As detailed in Equation~\ref{eq:encoder_loss}, we fix $\lambda_2=1$ and $\lambda_3=2$, and by adjusting the value of $\lambda_1$, we monitor changes in the model's performance. As depicted in Figure~\ref{fig: lambda_1}, with the increase of $\lambda_1$, the mAP consistently improves. However, for the MAE metric in direction-of-arrival (DoA), the performance gradually deteriorates. This presents a trade-off between the two metrics. Prioritizing classification, we opt for settings that yield the highest mAP. In subsequent experiments, we fix the hyperparameters at $\lambda_1=1250, \lambda_2=1, \lambda_3=2$.

\begin{figure}
\centering
\begin{minipage}[t]{.48\linewidth}
\includegraphics[trim={50 10 100 70},clip, width=\linewidth]{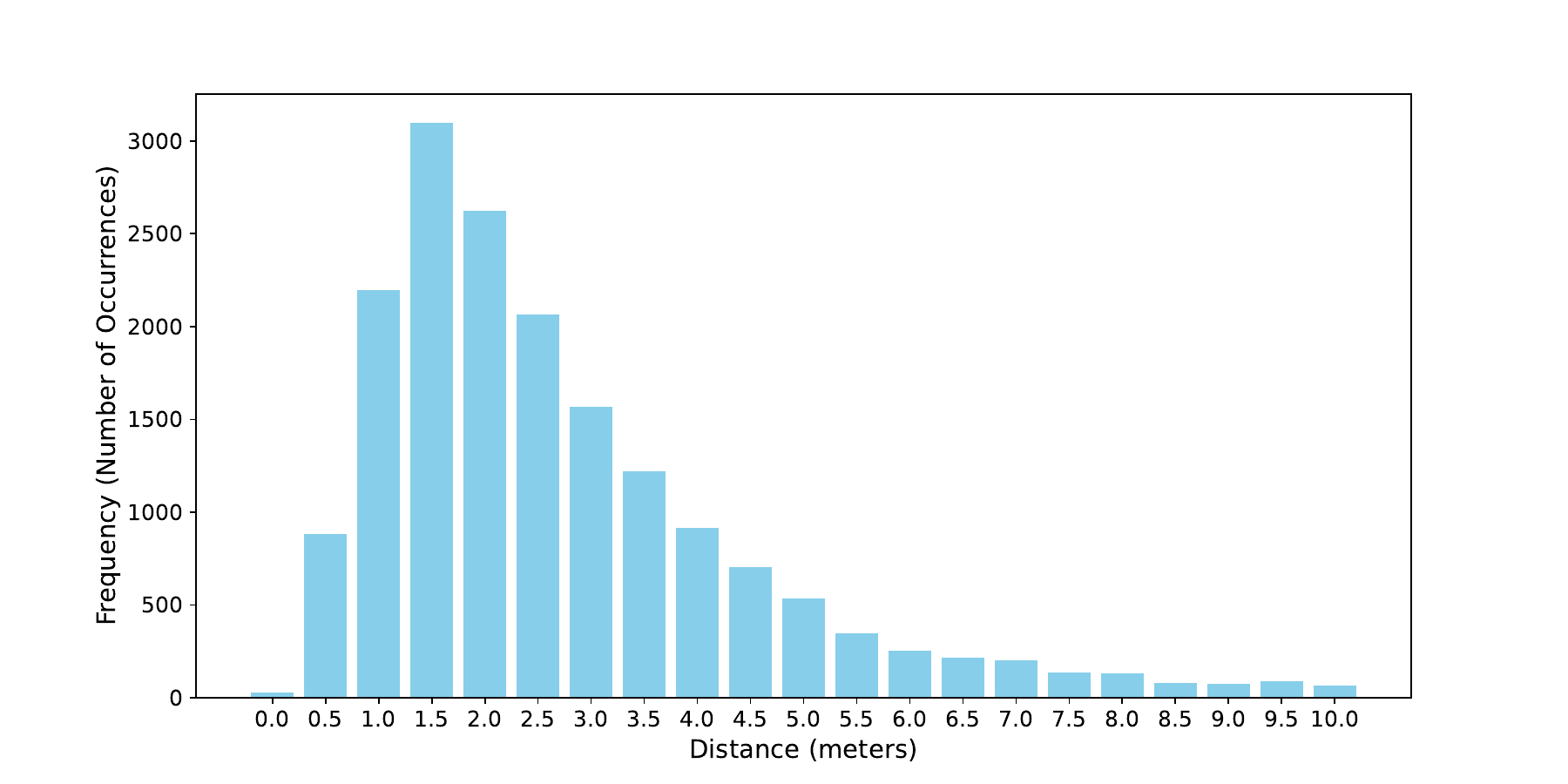} 
\vspace{-0.8cm}
\caption{Frequency distribution of training reverberations' distances ranging from 0 to 10 meters.}
\label{fig: distribution}
\end{minipage}
\hspace{0.4cm} 
\begin{minipage}[t]{.48\linewidth}
\vspace{-4cm}
\captionof{table}{Ablation study on monaural audio input: We employ our proposed architecture to train and infer under four distinct scenarios, with and without audio normalization and reverberation.}
\label{tab:normal_reverb}
\setlength\tabcolsep{4pt} 
\vspace{0.25cm}
\begin{tabular}{cccc}
\toprule
Normalization & Reverberation & mAP (\%) ↑ & DER (\%) ↓ \\
\midrule
\CROSS & \CROSS & \textcolor{gray}{52.64} & 49.49 \\
\CHECK & \CROSS & \textcolor{gray}{52.87} & 49.51 \\
\CROSS & \CHECK & 51.13 & 39.57 \\
\CHECK & \CHECK & 51.39 & \textbf{26.87} \\
\bottomrule
\end{tabular}
\end{minipage}
\end{figure}

\textbf{Audio Normalization and Reverberation.} In this ablation study, we assess the effects of audio normalization and reverberation on the encoder's performance, using monaural and fixing $\lambda_1, \lambda_2, \lambda_3=1250, 1, 2$ for simplicity, as outlined in Table~\ref{tab:normal_reverb}. The addition of reverberation indeed impacts the model's ability to detect sound events, as it makes the task more challenging without adding extra event information. A comparison between rows 1 and 2 with rows 3 and 4 indicates that removing reverberation significantly enhances mean Average Precision (mAP), increasing from 51.13\% to 52.64\%. However, it's noteworthy that audio without reverberation lacks spatial information, leading to a Distance Error Rate (DER) for distance perception that is nearly equivalent to random guessing. Based on the distance distribution described in Figure~\ref{fig: distribution}, random sampling yields approximately 67\% DER, and when sampling is limited to specific distances like 1.0, 1.5, and 2.0 meters, it results in around 53\% DER. The implementation of audio normalization further improves the encoder's performance in both classification and distance tasks. Comparing rows 3 and 4, the mAP increases slightly from 51.13\% to 51.39\%, while DER significantly drops from 39.57\% to 26.87\%, underscoring the benefits of audio normalization in our model. 

\section{\dataname~Generation}
\begin{figure}[!htb]
    \centering
    \includegraphics[trim={25 100 10 35},clip, scale=0.4]{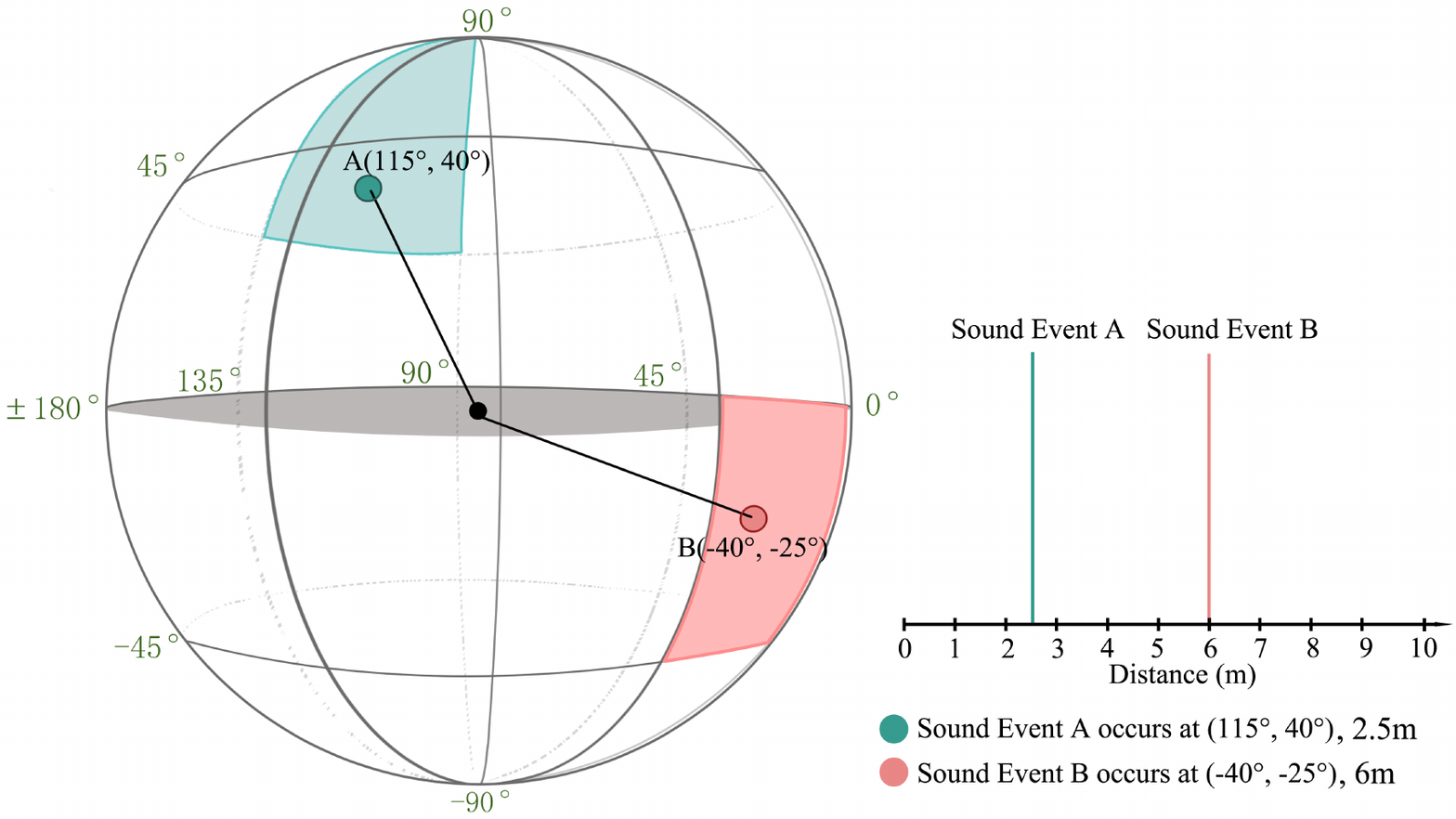}
    \caption{Sound events A and B occur in different directions and distances simultaneously.}\label{fig: 3dsphere}
\end{figure}
In the process of generating \dataname, we simulate spatial audio scenarios where multiple sound events occur simultaneously, each at distinct directions and distances from the receiver. For example, as depicted in Figure~\ref{fig: 3dsphere}, consider a situation where two sound sources, A and B, are active concurrently. Sound event A occurs at an azimuth of 115 degrees and an elevation of 40 degrees, situated 2.5 meters away from the receiver. In contrast, sound source B occurs at coordinates of -40 degrees azimuth and -25 degrees elevation, positioned 6 meters distant. It's important to note that in the actual dataset construction, the direction and distance parameters are often not as distinct as in this example. With these ground truth spatial parameters, we can then construct a diverse array of questions and answers concerning the spatial relationships between the two sound sources, as shown in Table~\ref{tab: data_overview}.

In supplement to Table 1, we provide details for each question type, including the initial prompt used and the corresponding answer template.

\begin{table}[!htp]
    \caption{Prompts and answer templates for \dataname}\label{tab: prompt_detail}
    \centering
    \footnotesize
    \begin{tabularx}{\textwidth}{l >{\raggedright\arraybackslash}X l}
    \toprule
       \textbf{Type} & \textbf{Prompt} & \textbf{Answer template} \\
    \midrule
        \textbf{A:} Detection & Identify the sound events in the audio clip. & sound event1; sound event2 \\
    \midrule
        \multirow{2}{*}{\textbf{B:} DP \& DoA} & Where is the audio clip coming from? & \multirow{2}{*}{\{d1\}, \{d2\}, \{d3\}; \{distance\}} \\
        & How would you describe the location of the \{sound event\}'s sound? \\
    \midrule
        \textbf{C:} Detection & Identify the sound events in the audio clip coming from the \{d1\}, \{d2\}, \{d3\}, approximately \{distance\} away. & sound event1; sound event2 \\ 
    \midrule
        \textbf{D:} DP \& DoA & Where is the sound of the \{sound event\} coming from? & \{d1\}, \{d2\}, \{d3\}; \{distance\}\\
    \midrule 
        \multirow{8}{*}{\textbf{E:} Reasoning} & Do the sound of \{sound event1\} and the sound of \{sound event2\} both come from your \{d\} side? & \multirow{8}{*}{Yes/No} \\
        & Can you hear the \{sound event1\}'s sound from the \{d1\} and the \{sound event2\}'s from the \{d2\}? & \\
        & Would you find the sound of \{sound event1\} on the \{d\} side of the sound of \{sound event2\}? \\
        & In terms of straight-line distance, does the sound of \{sound event1\} reach you from a closer point compared to the sound of \{sound event2\}? & \\
    \bottomrule
    \end{tabularx}
\end{table}

\section{Prompt Details}
For our spatial audio-based large language model \modelname, we designed a specific prompt to guide its responses to various spatial audio tasks. The prompt structure is as follows:
\begin{verbatim}
"Based on the audio you've heard, refer to the instruction and provide a response.\n\n
### Instruction:\n{instruction}\n\n### Response:"
\end{verbatim}

To form the input, we replace ``\{instruction\}" with our specific question, and then concatenate a fixed length of learnable query tokens at the beginning.

\section{Ablations of \modelname}\label{appendix: bat}
\begin{table*}[!htb]
    \centering
    \caption{The influence of different training curriculum on \modelname's reasoning ability.} \label{tab: bat_curriculum}
    \begin{threeparttable}
    \begin{tabular}{@{}ccccccc|c@{}}
\toprule
     \multirow{3}{*}{Model} & \multirow{3}{*}{\#Sources} & \multicolumn{3}{c}{Stage} & \multicolumn{3}{c}{Reasoning (BA~$\uparrow$)} \\ \cmidrule(lr){3-5} \cmidrule(lr){6-8}
     & & \uppercase\expandafter{\romannumeral1} & \uppercase\expandafter{\romannumeral2} & \uppercase\expandafter{\romannumeral3} & Direction & Distance & Avg. \\ 
\midrule
    Random & - & - & - & - & 50 & 50 & 50 \\ 
\midrule
    \multirow{7}{*}{\modelname} & \multirow{7}{*}{2} & \CHECK & \CROSS & \CROSS & 0 & 0 & 0 \\
    & & \CHECK & \CHECK & \CROSS & 0 & 0 & 0 \\
    & & \CROSS & \CROSS & \CHECK & 65.12 & 81.98 & 73.55 \\
    & & \CROSS & \CROSS & \CHECK\tnote{*} & 59.59 & 49.76 & 54.68 \\
    & & \CROSS & \CHECK & \CHECK\tnote{*} & 64.69 & 79.16 & 71.93 \\ 
    & & \CHECK & \CHECK & \CHECK\tnote{*} & 69.56 & 83.51 & 76.54 \\
    \cmidrule(lr){3-8}
    & & \CHECK & \CHECK & \CHECK & \textbf{69.77} & \textbf{84.01} & \textbf{76.89} \\
\bottomrule
\end{tabular}
 \begin{tablenotes}
\item[\raisebox{-2ex}{*}]\footnotesize{We removed question types C and D from \dataname.}
\end{tablenotes}
\end{threeparttable}
\end{table*}

In Table~\ref{tab: bat_curriculum}, we thoroughly examine the influence of different training curriculum on \modelname's reasoning ability. Initially, the model encounters difficulties with reasoning tasks, particularly when facing unfamiliar reasoning-style questions, as evidenced by its inability to produce accurate binary (Yes/No) responses. A one-stage version of \modelname~surprisingly achieves a respectable overall Binary Accuracy (BA) of 73.55\% in reasoning tasks. However, when we remove question types C and D, which are related to sound source separation, from the one-stage training, the model's reasoning ability drops significantly, exhibiting near-random performance with a BA of only 54.68\%. Intriguingly, when \modelname~is first trained on sound source separation tasks before proceeding to the final reasoning stage, it retains its ability to reason effectively, even in the absence of separation-related questions during the final stage. 
The last row of Table~\ref{tab: bat_curriculum} showcases the optimal results achieved through our three-stage curriculum, demonstrating the importance of a comprehensive training approach in equipping \modelname~with robust reasoning skills.

\begin{table}[!htb]
    \centering
    \begin{minipage}{.45\linewidth}
    \setlength\tabcolsep{3.5pt}
      \centering
      \caption{Impact of different tokens from \textsc{Spatial-AST} on One-stage \modelname's reasoning ability.} \label{tab: tokens}
      \begin{tabular}{@{}cccc|c@{}}
      \toprule
       \multirow{3}{*}{CLS tokens} & \multirow{3}{*}{Audio Tokens} & \multicolumn{3}{c}{Reasoning (BA~$\uparrow$)} \\ \cmidrule(lr){3-5}
       & & Direction & Distance & Avg. \\ 
      \midrule
      \CHECK & \CROSS & 60.32 & 72.92 & 66.62 \\
      \CROSS & \CHECK & 61.70 & 78.95 & 70.33 \\
      \CHECK & \CHECK & \textbf{65.12} & \textbf{81.98} & \textbf{73.55} \\
      \bottomrule
      \end{tabular}
    \end{minipage}
    \begin{minipage}{.5\linewidth}
    \centering
    \caption{Impact of learnable query size on reasoning accuracy.}\label{tab: query_size}
      \begin{tabular}{ccc|c}
      \toprule
       \multirow{3}{*}{\shortstack[3]{Learnable \\ Query}} & \multicolumn{3}{c}{Reasoning (BA~$\uparrow$)} \\ \cmidrule(lr){2-4}
       & Direction & Distance & Avg. \\ 
      \midrule
    32 & 61.42 & 83.49 & 72.45 \\
    64 & \textbf{69.77} & \textbf{84.01} & \textbf{76.89} \\
      \bottomrule
      \end{tabular}
    \end{minipage} 
\end{table}

Table~\ref{tab: tokens} demonstrates that the combination of both CLS and audio tokens yields the best performance in reasoning tasks. However, it is noteworthy that using only three CLS tokens as a representation for spatial audio also results in impressive performance. As per Table~\ref{tab: query_size}, a larger learnable query size in the projection module correlates with improved reasoning efficacy.



\end{document}